# Femtosecond pulsed laser micro-machining of glass substrates with application to microfluidic devices

M. S. Giridhar, Kibyung Seong, Axel Schülzgen, Pramod Khulbe,
Nasser Peyghambarian, and Masud Mansuripur

Optical Sciences Center, University of Arizona, Tucson, Arizona 85721

**Abstract.** We describe a technique for surface and sub-surface micro-machining of glass substrates using tightly focused femto-second laser pulses at a wavelength of 1660 nm. Although silicate glass is normally transparent at this wavelength, the extremely high intensity of the focused beam causes multi-photon absorption, resulting in localized ablation of the glass substrate. Ablation is strictly confined to the vicinity of focus, leaving the rest of the substrate unaffected. We exploit this phenomenon to drill a micro-hole through a thin vertical wall that separates two adjacent pits machined by the same laser in a glass plate. A salient feature of pulsed laser micro-machining, therefore, is its ability to drill sub-surface tunnels and canals into glass substrates, a process that requires multiple steps in standard lithography. To demonstrate a potential application of this micro-machining technique, we have fabricated simple micro-fluidic structures on a glass plate. To prevent the evaporation of liquids in open micro-channels and micro-chambers thus fabricated requires a cover plate that seals the device by making point-to-point contact with the flat surface of the substrate. This point-to-point contacting is essential if the fluids are to remain confined within their various channels and chambers on the chip, without leaking into neighboring regions. Methods of protecting and sealing the micro-machined structures for microfluidic applications are also discussed.

*Keywords*: laser micro-machining, femtosecond pulsed laser, microfluidics

**1. Introduction.** The availability of widely tunable femto-second laser sources in recent years has given rise to the development of many novel micro-fabrication and micro-machining techniques. Direct writing in bulk glass has been demonstrated, and optical components such as waveguides and gratings have been fabricated [1-3]. Most previous studies used pulses of ~100 fs duration at λ ~ 800nm; micro-machining was then performed by tightly focusing the pulse train beneath the surface of a glass plate or substrate. Ordinary glass shows no linear absorption at the incident wavelength; however, due to extremely high intensities at the focal point, multi-photon absorption occurs, resulting in the deposition of a large amount of energy in an extremely small volume [4-7]. This process modifies the local refractive index of the material and is exploited for fabricating waveguides and other light-manipulating devices. Recently, drilling a hole through bulk glass has also been demonstrated [8].

This paper describes another application of femtosecond laser writing based on the ablation of substrate material at or below the surface. Instead of focusing the pulsed laser beam in the bulk of the material – as is common in the fabrication of optical waveguides and other micro-devices – we typically focus the beam at or near the surface of the substrate, which results in localized ablation. By appropriately translating the substrate under computer control, we have fabricated microfluidic pathways and chambers 10-100μm wide and 5-50μm deep on several glass plates. Wide features are written by rastering the beam over the desired zone, while the depth and profile of written structures are controlled by adjusting the laser power and/or by multiple scanning of a given region.

In parallel with the development of sophisticated micro-fabrication techniques, there has been a flurry of activity relating to the so-called "biochip" devices for applications such as rapid on-site gene sequencing, biochemical sensing, biologically-inspired computing, and biological memories. To be successful, these devices must be compact, mass-producible, flexible in their application, robust, and cost efficient. Rapid prototyping techniques play a key role in the development of such devices. A common requirement for all these devices is fast and efficient transport of biomolecules and other chemicals in the liquid phase from one location to the other. Microfluidic pathways thus play a crucial role in the functioning of biochip devices. The small volume of liquids involved means that even nominal evaporation rates can cause rapid loss of the liquids. To prevent evaporative losses, we describe a method of covering and sealing our µ-machined structures with transparent sheets of polydimethyl-siloxane (PDMS). In recent years, PDMS has been widely used in microfluidic device fabrication because of its ease of use and compatibility with biological materials [9-12].

**2. Experimental**. Our set up is shown in Fig.1. A train of femto-second pulses is obtained from an Optical Parametric Amplifier (OPA) of a *Spectra Physics* laser system. The OPA is tuned to produce pulses of ~130fs duration at a repetition rate of 1 KHz. The wavelength of the light is $\lambda = 1660$ nm and the maximum energy per pulse is 20µJ. When focused via a 0.25NA infrared microscope objective, the laser beam produces a diffraction-limited spot of diameter $\sim \lambda / NA \sim$ 7µm at a distance of 5mm from the front facet of the objective lens. (The beam fills the entrance pupil of the objective.) For incident pulse energies near the maximum, the laser is intense enough to produce a bright luminescent spot caused by a breakdown of the air within the focal region.

Neutral density filters are introduced into the beam path to attenuate the beam's intensity. The sample is mounted on a translation stage that can be moved under computer control. The computer also controls a shutter in the beam path to switch the beam on and off at desired locations during the writing process. Before commencing a run, the sample surface is made perpendicular to the incident beam by ensuring that the retro-reflected light passes through suitably positioned irises along the beam path.

**3. Micro-machining of surface channels**. Micro-channels are fabricated by focusing the beam directly on the top surface of a glass substrate, then translating the substrate at ~20µm/s. To observe the cross-sectional profile of these channels, we extend the writing all the way to one edge of the substrate. Scanning Electron Micrographs (SEM) and optical microscope images of typical channels are shown in Fig.2. The cross-sectional view in Fig.2(b) reveals the depth profile of our multi-scan channels. The channel depth can be reduced by attenuating the incident beam. Wider channels can be written by scanning the beam along parallel, overlapping lines by keeping the separation between adjacent scans below the 7µm diameter of the focused spot. The roughness of the channel sidewalls is readily visible in the optical micrographs of Figs.2(c, d).

**4. Chambers connected via a sub-surface tunnel**. A unique feature of laser µ-machining based on ablation caused by multiphoton absorption is that the ablation is tightly confined to the focal volume. Regions surrounding the focal volume are almost completely transparent to the incident radiation and remain untouched by the laser beam. We have exploited this effect and drilled a µ-hole through the partition wall between two adjacent µ-chambers, as shown schematically in Fig.3.

The substrate used in this work was a regular glass microscope slide. In the first step, we polished one edge of the slide (using a Logitech PM5 polishing machine) at 25 rpm for one hour. In the second step, we machined three chambers by scanning the top surface of the substrate with the focused laser beam, as shown in Fig. 4. One of the chambers is used for calibration purposes while the other two, connected via a µ-hole at the center of their partition wall, are used in microfluidic experiments. Each chamber is ~190μm on the side and ~90μm deep. Two critical parameters for controlled drilling of the μ-hole are thickness of the partition wall (between adjacent chambers) and thickness of the outer wall (near the edge of the substrate). The outer wall, which functions as an optical window, must be smooth and homogeneous if the laser beam is to pass through it unperturbed on its way to focus at the partition wall. When the outer wall was too thin, it was damaged during drilling of the μ-hole (exterior surface roughened) and eventually collapsed. Similarly, when the partition wall was too thin (for a given laser power), it could not withstand drilling and was demolished in the process. On the other hand, thick partition walls prevented the μ-hole from penetrating the entire thickness of the wall. For the cone of light emerging from the objective, the $190 \times 190 \mu m^2$ size of the μ-chamber was another factor that needed consideration while optimizing the drilling process. We found the optimum thickness of the outer wall to be ~50μm, while that of the partition wall was ~25μm. Optimum conditions for writing the chambers were: translation speed ~50μm/s, time-averaged laser power ~18mW, steps in the XY plane ~ 4μm, and steps in the depth direction ~15μm.

In the third step, the sample was turned over by 90° and the beam was focused at the center of the partition wall between the two adjacent chambers; see Fig. 3. The μ-hole was drilled in the partition wall with the beam passing through the (polished) outer wall of the chamber. After rotating the sample we located the chambers by attenuating the beam to non-machining intensity levels and observing the strong scattering that occurred when the beam hit the edges of the chambers. Again an orthogonal rastering procedure was used to drill the hole in the partition wall (translation speed ~1-5μm/s in the XY plane of the scan, with 2μm steps in the depth-direction). To prevent the debris from depositing in the chambers or surrounding areas, the substrate was exposed to a mild stream of dry nitrogen during the entire writing process. SEM and optical images of the chambers with their perforated wall are shown in Fig. 4. The μ-hole diameter is ~40μm on the front side and ~15μm on the rear side. In Fig. 4(a) the front surface of the outer wall is seen to have been slightly damaged by exposure to the laser beam during drilling. Figure 4(f) shows a cross-sectional view of the μ-hole as seen through an optical microscope focused under the top surface of the partition wall.

Apart from being a good demonstration of the proposed micro-machining technique, the structure we have thus produced finds application in bio-chip devices that rely on DNA (or other polymer) translocation through nano-pores. The two adjacent chambers act as the *cis* and *trans* chambers, and the lipid bilayer membrane (host to proteinaceous nano-pore) can be painted over the opening of the μ-hole. The conical shape of the μ-hole may in fact prove to be advantageous in such experiments as it might enhance the pressure gradients, thus facilitating the passage of DNA molecules through the nano-pore.

**5. Microfluidic structures**. A potential application of femtosecond laser μ-machining is direct fabrication of microfluidic pathways on glass substrates. This method can create larger and/or deeper surface and sub-surface features than is feasible via conventional lithographic techniques. In the remainder of this paper we describe methods of fabricating sealed micro-fluidic pathways – as required, for instance, in practical bio-chip devices – and examine the microfluidic behavior inside properly sealed chambers.

We developed two methods of covering and sealing the μ-machined chambers and channels produced by femtosecond laser writing. Attempts to cover these devices with flat glass plates proved futile as we failed to achieve point-to-point contact between the substrate and the cover plate due to the roughness near the edges of machined µ-structures. We found the liquid under the cover plate to leak out of one channel and enter adjacent channels, no matter how hard we tried to bring the two pieces of glass into perfect contact. On the other hand, PDMS material proved to be ideal for covering and sealing these microfluidic structures. In one approach, we baked a sheet of PDMS and placed it directly on the top surface of the µ-machined sample. Application of mild, uniform pressure over the PDMS sheet was sufficient to cause point-to-point adhesion to the surface of the substrate. In the second approach, we filled the machined µ-structures with photoresist (as a sacrificial filler material), then polished the surface and proceeded to spin-coat liquid PDMS on the polished substrate. After baking the sample (to solidify its PDMS cover layer), the residual resist was dissolved and removed from channels and chambers without damaging the cover sheet (or adversely affecting the point-to-point contact between the two surfaces). Detailed descriptions of these techniques are given in the following subsections.

**5.1. Covering µ-channels with preformed PDMS sheets**. PDMS (10:1 by volume of base and cross-linking agent, Sylgard 184, Dow Chemicals) was spin-coated onto $2 \times 2 cm^2$ polished Teflon substrates. We used a one-step spinning process at 1000rpm for 60 seconds. Teflon was used as a substrate as it allowed easy detachment of the polymer sheet. After spin coating the Teflon substrate was cured at 60°C for one hour, then a clean razor blade was used to cut out a section of the baked PDMS layer. The PDMS sheet thus produced was placed over the µ-channels, leaving ~1mm from one end of the sample exposed to facilitate µ-pipette insertion for introducing liquids into the channels. Micro-pipettes with outer tip diameters less than 10µm were drawn from glass capillary tubes using a commercial pipette puller (Narishige PD-5, Japan). The pipettes were mounted on XYZ stages and connected via flexible tubings to syringes or to a micro-injector. The pipette tips were made with a slow taper, making them sufficiently flexible to help in aligning and inserting the tips into the µ-channels. Figure 5 shows the process of inserting a pipette tip into a µ-channel beneath the PDMS layer. An air bubble is seen to develop under the PDMS film in the vicinity of the insertion point because the small angle between the pipette and the plane of the substrate tends to lift the PDMS layer off the glass substrate. The size of the air gap thus limits the range of allowed separations between adjacent µ-channels in this configuration. Figure 6 shows an array of covered µ-channels that connect a µ-chamber (also covered) to the edge of the host substrate. In Fig. 6(d) water injected into one of the channels is seen to fill the µ-chamber, then return through the other channels.

**5.2. Spin-coating PDMS films on µ-machined structures**. One drawback of laying down preformed PDMS sheets on µ-machined structures is that very often air bubbles get trapped between the cover sheet and the substrate. If these bubbles are large enough to span the distance between adjacent channels, they will cause fluidic cross-talk between the channels. This problem is specially severe for large-area samples. Spin-coating the PDMS layer directly on µ-channels filled with sacrificial photoresist is one approach to solving the aforementioned problem. We fabricated surface channels ~1.5mm-long, running off the edge of a microscope slide (i.e., glass substrate). The machined slides were sonicated in acetone for ~30 minutes to remove any machining debris. Positive photoresist, Shipley S1813, was then spin-coated on the sample at 900rpm for 60 seconds. Subsequently, the slide was baked at 110°C for two minutes. The resist formed a hard, transparent, ~50µm-thick coating over the entire substrate, filling the channels uniformly. The coated facet was then polished using a Logitech PM5 polishing machine. A fine-finish polyutherene polishing wheel rotating at 25rpm was used with a cerium oxide slurry (particle size ~0.5µm). During polishing the sample was frequently checked to avoid over-polishing. (The resist had an orange tint to it, and it was possible to determine by visual inspection the point at which the resist was completely removed from the surface.) At this point the top surface of the resist-filled channels was flush with the substrate surface. Figure 7(a) is an optical micrograph of empty channels (before being filled with resist), while Fig. 7(b) shows the filled channels after polishing.

A PDMS solution, prepared as described in section 5.1, was spin-coated on the polished substrate at 900rpm for 60 seconds. This resulted, after curing at 60°C for one hour, in a 20µm-thick PDMS film over the entire substrate. A clean razor blade was used to cut off a narrow strip of PDMS along the edge of the sample, perpendicular to the channels. This produced a step, exposing ~100µm of the resist-filled channels. The sample was then placed vertically in a dish of continuously stirred acetone with the PDMS step fully submerged. Acetone (which does not interact with PDMS) dissolved and removed the residual resist material in the µ-channels beneath the cover layer. After ~20 minutes of acetone wash, the samples were coated with gold for observation under a Scanning Electron Microscope (SEM).

**6. Flow through µ-hole in sealed microfluidic device**. To observe the passage of liquids through the µ-hole that joins two of the three chambers of Fig. 4, we covered the sample with a 20µm-thick PDMS sheet. The preformed PDMS film was peeled off from its Teflon backing plate and placed over the µ-chambers without exerting any force. To enhance the adhesion of the PDMS layer to the substrate, the covered structure was baked for an additional three minutes at 60°C. We then inserted, by pushing from the top, two µ-pipettes (10µm diameter) into the two adjacent chambers, one for injecting the deionized water, the other for extracting the trapped air; see Fig. 8. We designed the µ-pipette tips for sufficient stiffness by controlling the force applied during their pulling as well as the temperature used for their heating and softening. Also, for effective puncturing of the PDMS layer, we adjusted the angle between the µ-pipette tip and the plane of the PDMS sheet at ~45°. A micro-injector was then used to fill one chamber with water using 60ms, 5 psi pressure pulses. (For effective circulation, it is possible to use a vacuum pump to apply negative pressure to the pipette through which air is intended to depart, although, strictly speaking, this is not necessary.) We were thus able to observe (under a microscope) the flow of the liquid through the µ-hole and the subsequent filling of the adjacent chamber. A sequence of the observed events is shown in Fig.8.

**7. Concluding remarks**. High-power femtosecond laser pulses can be used to create micro-structures at and below the surface of a glass substrate. For microfluidic applications, one can cover these structures with a PDMS sheet to protect the liquids inside the various channels and chambers, and also to seal the device so as to prevent the leakage and mixing of the contents of the various compartments.

**Acknowledgement**. The authors are grateful to Professor Raphael Gruener and Dr. Stephen Kuebler of the University of Arizona for illuminating discussions. This work has been supported in part by the *National Science Foundation* STC Program, under Agreement No. DMR-0120967.

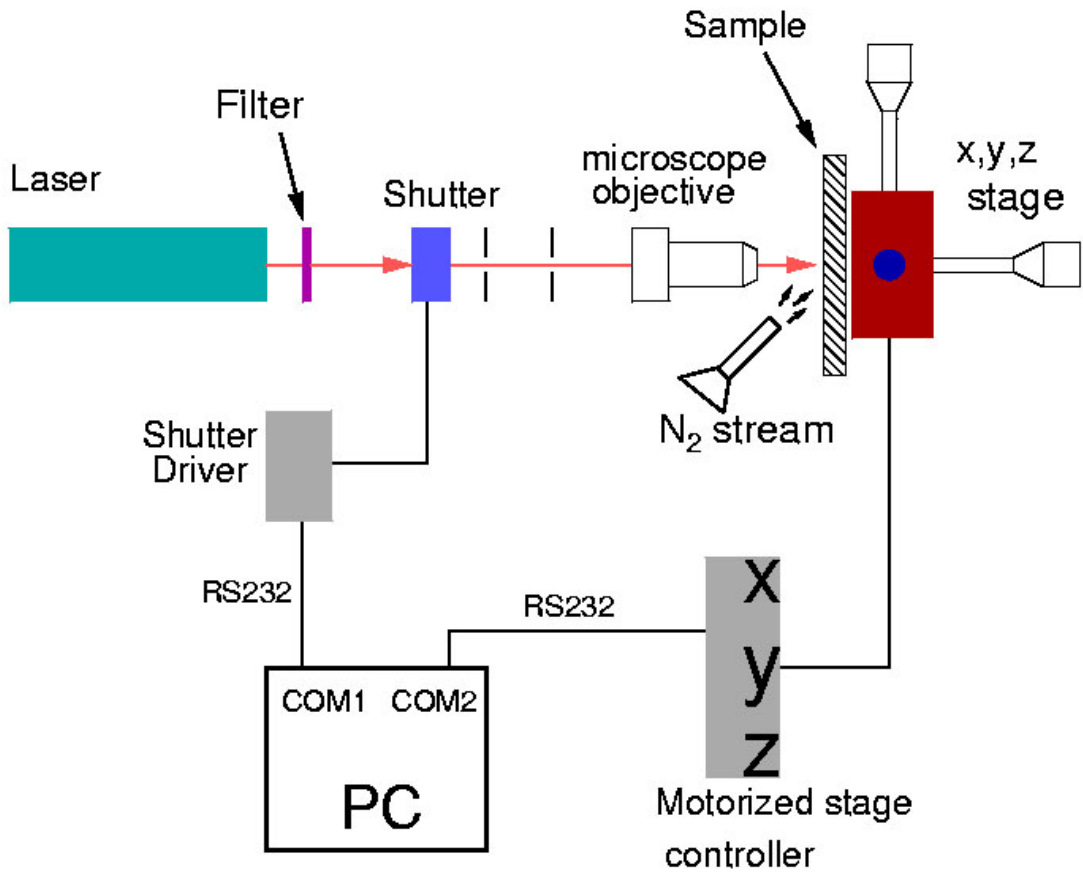

**Figure 1**. Diagram of the micro-machining experimental setup. The femtosecond pulsed laser is focused on the glass substrate via a microscope objective. The sample, mounted on a computer-controlled XYZ positioner, is moved in small steps along the X, Y, and Z-axes.

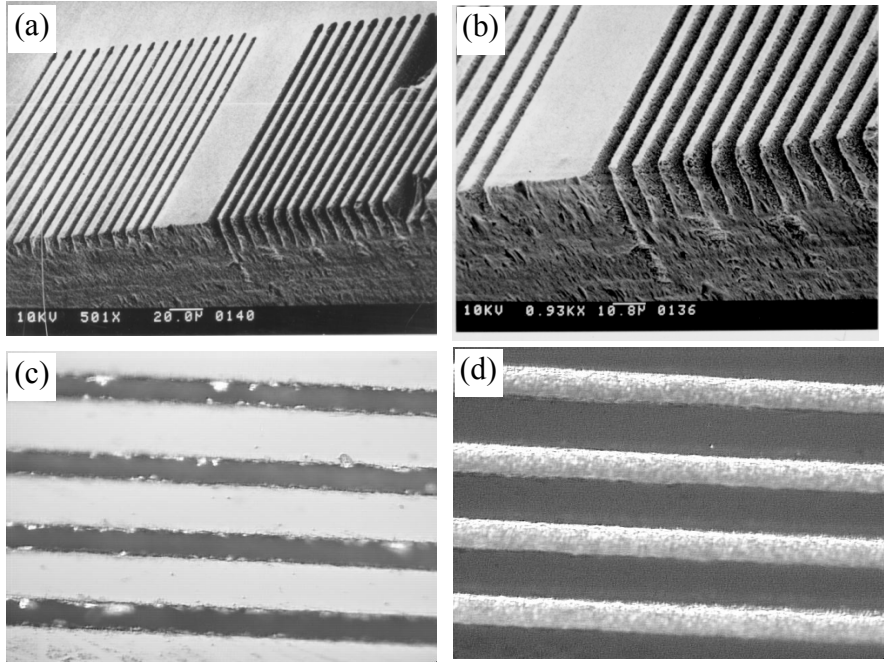

**Figure 2**. Various images of µ-machined channels on glass substrate. (a) SEM image showing parallel channels written near the edge of a glass slide; the group on the right-hand side is written with a larger dose of laser pulses. The channels are ~5µm wide and several microns deep. (b) Close-up view of the µ-channels as seen through an SEM focused on the edge of the substrate. (c) Optical micrograph obtained with top illumination; the µ-channels, seen as dark bands, are ~10µm wide, with a center-to-center spacing of ~30µm. (d) Optical micrograph of the same sample as in (c), obtained with illumination from below. The channels are seen as bright stripes on a dark background.

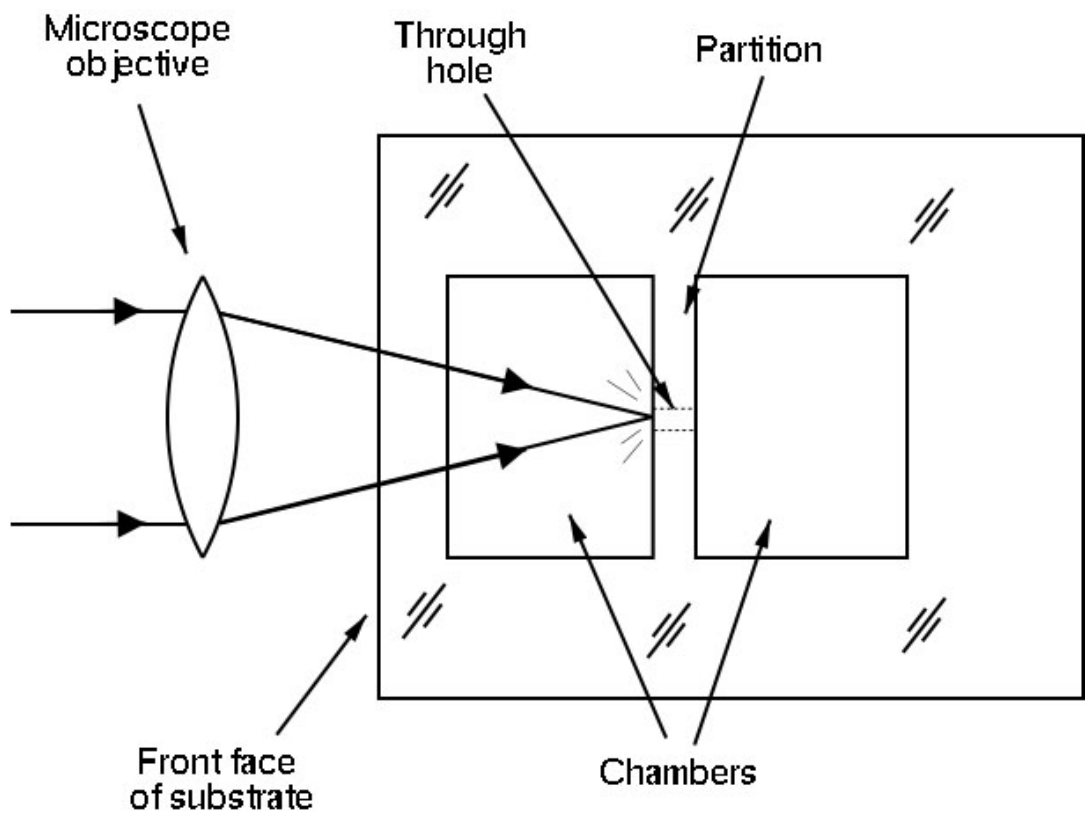

**Figure 3**. Diagram showing the focusing of the laser beam through a sidewall onto the partition wall between adjacent chambers.

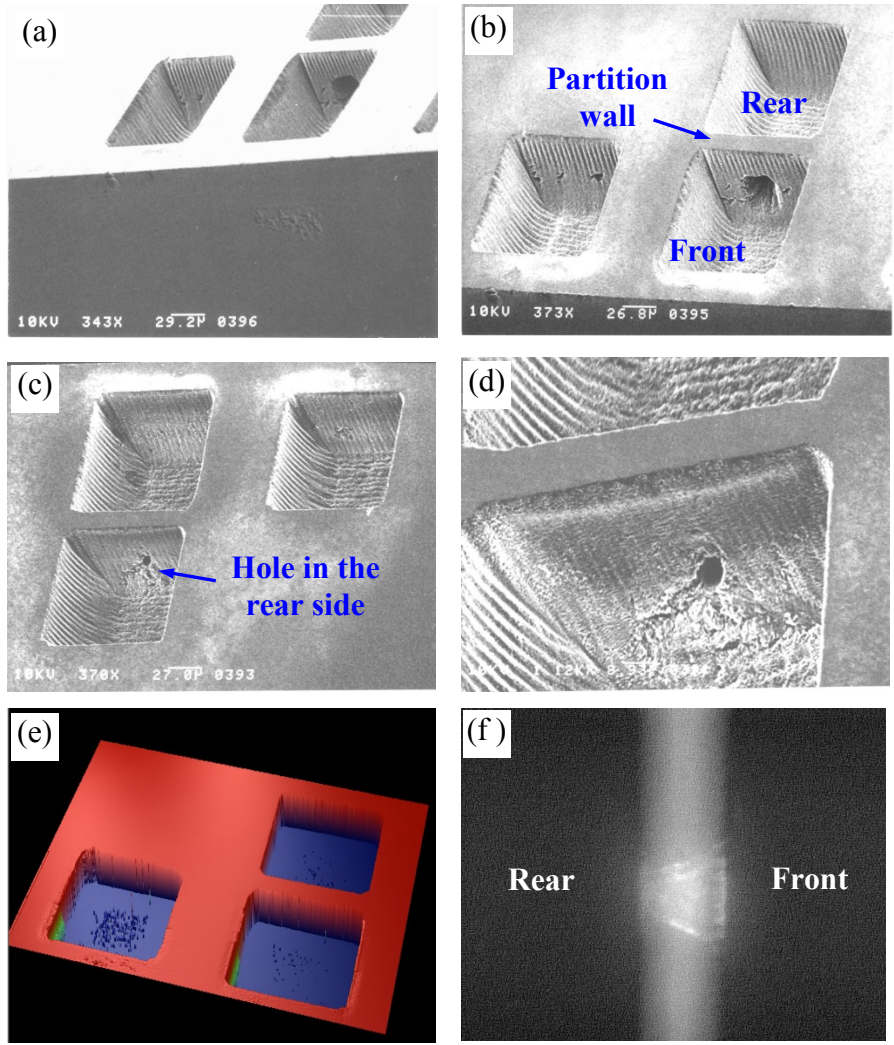

**Figure 4**. Micro-chambers written near the edge of a glass substrate. The three chambers are ~190 × 190 × 90 µm$^3$. The wall between the two chambers on the right-hand side has been perforated with a small, conical hole. (a) View from the side, showing slight damage to the front edge of the substrate through which the laser beam was focused to create the µ-hole. (b) Top view from the front side, showing the ~40µm diameter opening of the µ-hole. (c) Top view from the rear side, showing the ~15µm diameter of the tapered hole. (d) Magnified view of the tapered end of the hole. (e) Another view of the three chambers, obtained via an optical surface profiler. (f) Image of the µ-hole as seen through an optical microscope that looks straight at the top of the partition wall between two chambers; the tapered shape of the sub-surface hole is clearly visible in this image.

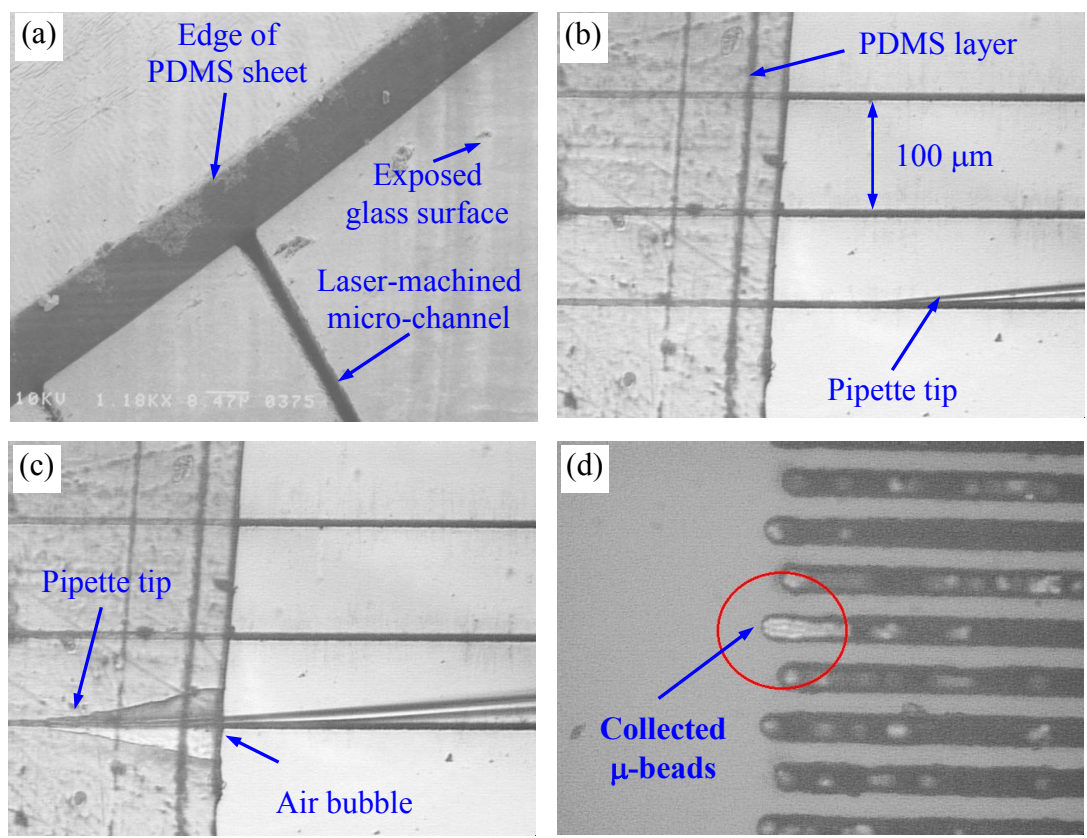

**Figure 5**. PDMS-covered µ-channels on a glass substrate. (a) The PDMS layer is on the left-hand side, covering part of a laser-machined µ-channel; the wide, dark diagonal band is the vertical edge of the PDMS layer. (b) An array of µ-channels is partially covered by a PDMS sheet, and a pipette tip is introduced into one of the channels. (c) Pushing the pipette tip under the PDMS sheet produces a small air bubble where the flexible tip (guided by the µ-channel) enters the covered region. (d) Photomicrograph of the end zone of an array of µ-channels, covered with a PDMS sheet and filled with water. The small white particles floating in the channels are 2µm-diameter µ-beads. These µ-beads were injected into the channels through a pipette, then guided to the end-zone by means of a focused laser beam (optical tweezers).

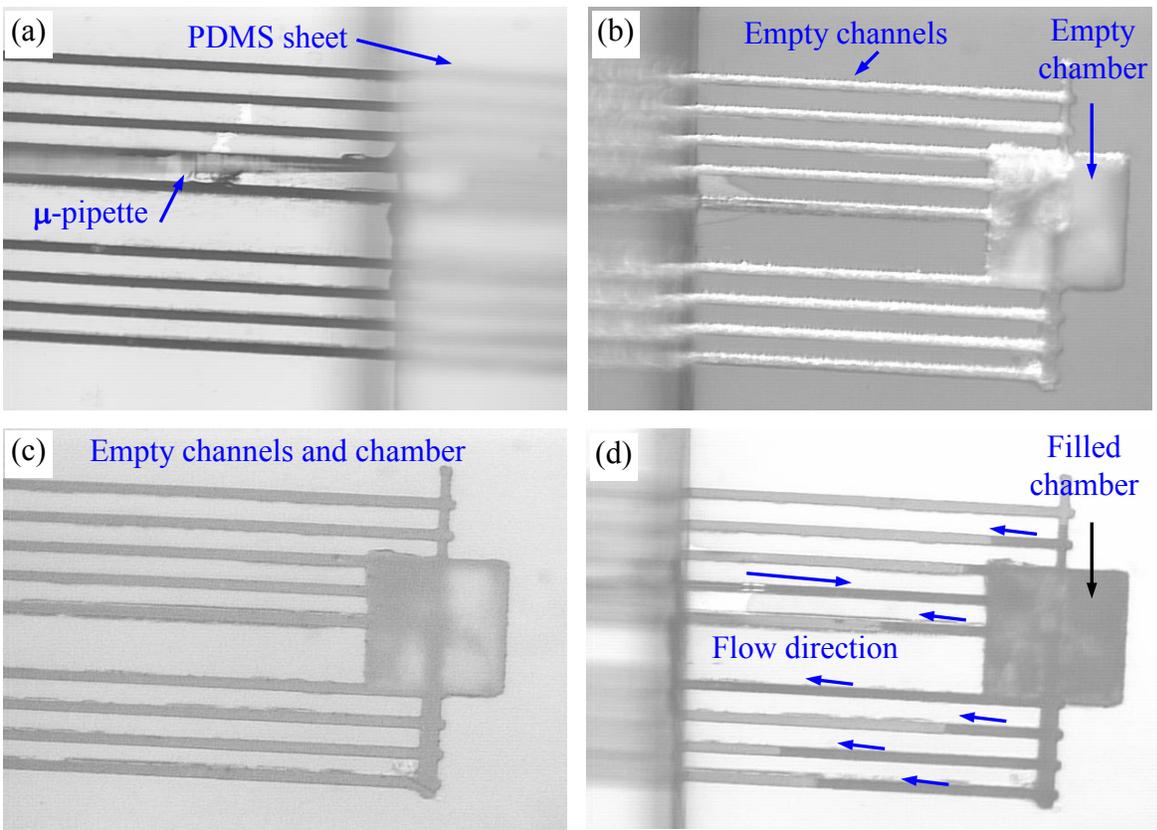

**Figure 6**. A $100 \times 100 \times 50\mu m^3$ chamber connected via nine μ-channels (width ~10μm, depth ~5μm) to the edge of the host glass slide. The chamber and parts of the channels are covered with a PDMS sheet. The chamber is subsequently filled with water injected into one of the channels. (a) Photo-micrograph of the bare section of the channels; a part of the (out-of-focus) PDMS sheet is also visible on the right-hand side. (b) Covered chamber and connecting channels as seen through the PDMS layer. The bare sections of the channels (seen on the left-hand side) are now out of focus. The picture, taken by illuminating the sample from below, shows the roughness of the channel walls and the uneven nature of the machined μ-chamber. (c) Same as (b), but illumination is from the top. (d) A μ-injector sends water through a pipette inserted into the fourth channel from the top of the picture. The presence of liquid in a region (channel or chamber) makes it appear darker than the empty regions. The liquid has now filled the chamber and is returning through several of the remaining channels. In particular, the fourth channel from the bottom is filled all the way to the edge of the PDMS layer.

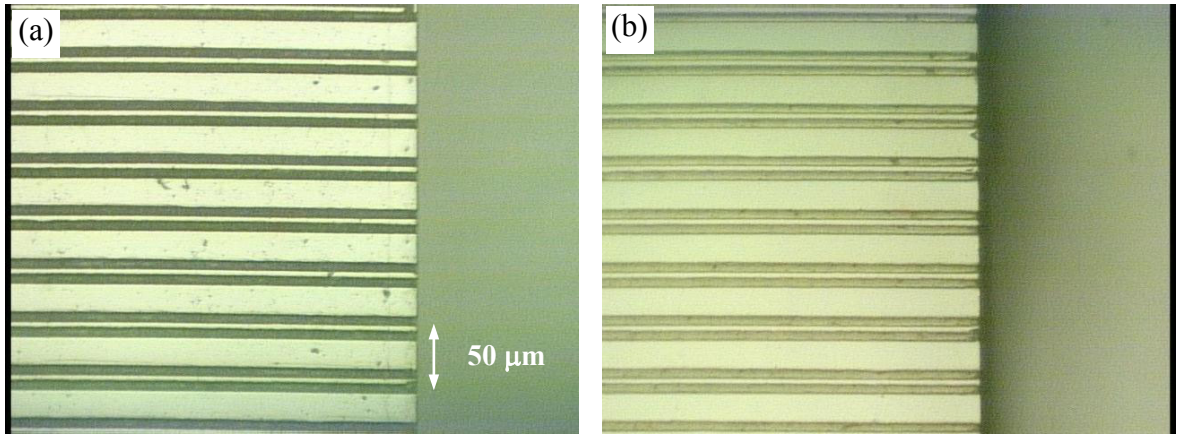

**Figure 7**. (a) Micro-machined parallel channels on a glass substrate. (b) Same channels after being filled with photoresist, then polished flush with the glass surface to remove debris and roughness in the areas between adjacent channels. Such polished substrates are covered with a PDMS sheet to create perfect (point-to-point) contact between the two surfaces. The resist is subsequently dissolved in acetone and removed.

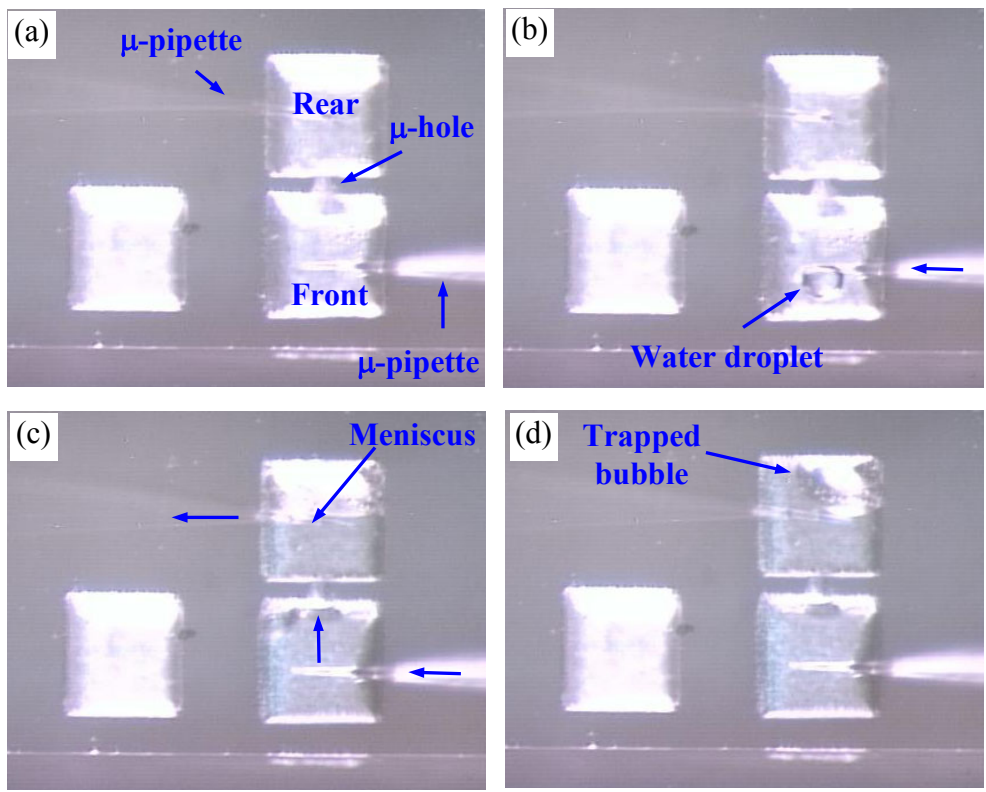

**Figure 8**. Water-filling procedure for the μ-chambers of Fig. 4, now covered with a 20μm-thick PDMS sheet. (a) Two μ-pipettes, each having ~20μm diameter at their tapered ends, approach and puncture the top surface of the PDMS layer. (b) Deionized water is injected by a μ-injector through one pipette, while the air from the adjacent chamber escapes through the other. (c) Water is seen to flow between the chambers through the μ-hole in the partition wall. (d) Aside from a small trapped air bubble, the second chamber is filled with water. In time the bubble escapes through the second pipette, and both chambers become completely filled.